\documentclass[10pt,letterpaper]{article}

\usepackage[aboveskip=1pt,labelfont=bf,labelsep=period,justification=raggedright,singlelinecheck=off]{caption}
\usepackage[top=0.85in,left=2.75in,footskip=0.75in]{geometry}
\usepackage{lastpage,fancyhdr,graphicx}
\usepackage{textcomp,marvosym}
\usepackage{amsmath,amssymb}
\usepackage{nameref,hyperref}
\usepackage[utf8]{inputenc}
\usepackage{changepage}
\usepackage[right]{lineno}
\usepackage{microtype}
\usepackage{epstopdf}
\usepackage{rotating}
\usepackage{fixltx2e}
\usepackage{cite}

\fancyheadoffset[L]{2.25in}
\fancyfootoffset[L]{2.25in}
\makeatletter
\renewcommand{\@biblabel}[1]{\quad#1.}
\makeatother

\raggedright
\DisableLigatures[f]{encoding = *, family = * }

\date{}

\usepackage[usenames,dvipsnames]{xcolor}
\usepackage{wasysym}
\usepackage{xspace}
\usepackage{helvet}
\usepackage{tikz}

\DeclareGraphicsExtensions{.pdf,.eps,.png,.jpg}

\newcommand{\resref}[0]{Results\xspace}
\newcommand{\metref}[0]{Methods\xspace}

\newcommand{\figref}[1]{Fig.~\ref{fig:#1}\xspace}
\newcommand{\tblref}[1]{Table~\ref{tbl:#1}\xspace} 
\renewcommand{\eqref}[1]{Eq.~(\ref{eq:#1})\xspace}
\newcommand{\figsref}[2]{Figs.~\ref{fig:#1} and~\ref{fig:#2}\xspace}

\newcommand{\eqsref}[2]{Eqs.~(\ref{eq:#1}) and~(\ref{eq:#2})\xspace}

\definecolor{orange}{RGB}{232, 153, 42}
\definecolor{light}{RGB}{225, 180, 119}
\definecolor{blue}{RGB}{19, 103, 177}

\begin{document} \vspace*{0.35in}

\begin{flushleft}
	{\Large\textbf\newline{Clustering Scientific Publications Based on Citation Relations: A Systematic Comparison of Different Methods}} \newline \\
	Lovro \v{S}ubelj\textsuperscript{1,*}, Nees Jan van Eck\textsuperscript{2}, Ludo Waltman\textsuperscript{2} \\ \bigskip
	\bf{1} University of Ljubljana, Faculty of Computer and Information Science, Ljubljana, Slovenia \\
	\bf{2} Leiden University, Centre for Science and Technology Studies, Leiden, Netherlands \\ \bigskip
	* lovro.subelj@fri.uni-lj.si
\end{flushleft}

% % % % % % % % % % % % % % % % % % % % % % % % % % 
%
%					ABSTRACT
%
% % % % % % % % % % % % % % % % % % % % % % % % % %

\section*{Abstract}

Clustering methods are applied regularly in the bibliometric literature to identify research areas or scientific fields. These methods are for instance used to group publications into clusters based on their relations in a citation network. In the network science literature, many clustering methods, often referred to as graph partitioning or community detection techniques, have been developed. Focusing on the problem of clustering the publications in a citation network, we present a systematic comparison of the performance of a large number of these clustering methods. Using a number of different citation networks, some of them relatively small and others very large, we extensively study the statistical properties of the results provided by different methods. In addition, we also carry out an expert-based assessment of the results produced by different methods. The expert-based assessment focuses on publications in the field of scientometrics. Our findings seem to indicate that there is a trade-off between different properties that may be considered desirable for a good clustering of publications. Overall, map equation methods appear to perform best in our analysis, suggesting that these methods deserve more attention from the bibliometric community.

\section*{Introduction}

There is an extensive literature on the topic of graph partitioning and community detection in networks~\cite{For10}. This literature studies methods for partitioning the nodes in a network into a number of groups, often referred to as communities or clusters. The general idea is that nodes belonging to the same cluster should be relatively strongly connected to each other, while nodes belonging to different clusters should be only weakly connected.

Which methods for graph partitioning and community detection perform best in practice? The literature does not provide a clear answer to this question, and if the question can be answered at all, then most likely the answer will be dependent on the type of network that is being studied and on the type of partitioning that one is interested in.

In this paper, we therefore address the above question in one specific context. We are interested in grouping scientific publications into clusters and we expect each cluster to represent a set of publications that are topically related to each other. Clustering scientific publications is a problem that has received a lot of attention in the bibliometric literature. In this literature, publications have for instance been clustered based on co-occurring words in titles, abstracts, or full text~\cite{BNDKPBSSMB11,JLGD06}, based on co-citation or bibliographic coupling relations~\cite{BK10,Jar07,SG74}, and sometimes even based on a combination of different types of relations~\cite{BK10,JGD08,Sma97,WVN10}. Following Waltman and Van Eck~\cite{WV12} and Boyack and Klavans~\cite{BK14,KB16}, our interest in this paper is in clustering publications based on direct citation relations. Direct citation relations are of special interest because they allow large sets of publications to be clustered in an efficient way. Waltman and Van Eck for instance cluster ten million publications from the period 2001-2010 based on about hundred million citation relations between these publications. In this way, they obtain a highly detailed classification system of scientific literature covering all fields of science.

The analysis presented in this paper focuses on systematically comparing the performance of a large number of clustering methods when applied to the problem of clustering scientific publications based on citation relations. The following clustering methods are included in the analysis: spectral methods~\cite{KK98a,DGK07}, modularity optimization~\cite{CNM04,BGLL08,RN11b,WV13}, map equation methods~\cite{RB08,RB11b}, matrix factorization~\cite{YL13}, statistical methods~\cite{LRRF11}, link clustering~\cite{ABL10}, label propagation~\cite{RAK07,SB11d,SB11e,SB14a,Gre10}, random walks~\cite{PL05}, clique percolation~\cite{KKKS08} and expansion~\cite{LRMH10}, and selected other methods~\cite{CRGP12,YML14}. These are all methods that have been proposed during the past years in the literature on graph partitioning and community detection.

To evaluate the performance of the different clustering methods, we perform an in-depth analysis of the statistical properties of the clusterings obtained by each method. On the one hand we focus on general properties of the clusterings, but on the other hand we also consider a number of properties that are of special relevance in the context of citation networks of publications. However, to obtain a deep understanding of the differences between clustering methods, we believe that analyzing the statistical properties of clusterings is not sufficient. Understanding the differences between clustering methods also requires an expert-based assessment of different clusterings. This is a challenging task that involves a number of practical difficulties, but in this paper we nevertheless make an attempt to perform such an expert-based assessment. The expert-based assessment is performed for publications in the field of library and information science, focusing on the subfield of scientometrics.

This paper is organized as follows. We first discuss the data and methods included in our analysis. We then present the results of the analysis. We conclude the paper by providing a detailed discussion of our findings.

% % % % % % % % % % % % % % % % % % % % % % % % % % 
%
%					METHODS
%
% % % % % % % % % % % % % % % % % % % % % % % % % %

\section*{Methods}

Below we first discuss the citation networks of publications that we consider in our analysis. We then discuss the clustering methods included in the analysis. Finally, we discuss the criteria that we use for comparing the clustering methods. These criteria relate to the following four properties of a clustering method:
\begin{description}
	\item[Cluster sizes.] Ideally the differences in the size of clusters should not be too large. For instance, the largest cluster preferably should be no more than an order of magnitude larger than the smallest cluster.
	\item[Small clusters.] For practical purposes, it is usually inconvenient to have a large number of very small clusters. Therefore the number of very small clusters should be minimized as much as possible.
	\item[Clustering stability.] Running the same clustering method multiple times may yield different results (due to random elements in many clustering methods), but the results should be reasonably similar. Likewise, when small changes are made to a citation network, this should not have too much effect on the results of a clustering method.
	\item[Computing time.]  Preferably, a clustering method should be fast. Especially in applications to large citation networks the issue of computing time is of significant importance.
\end{description}

In addition to the above four properties, a fifth property for comparing clustering methods is the intuitive sensibility of the results provided by a method. Experts should be able to interpret the clusters obtained from a clustering method in terms of meaningful research topics. We do not evaluate this fifth property using quantitative criteria. Instead, our expert-based assessment of the results of different clustering methods is focused on this criterion.

\paragraph{Citation networks of scientific publications.} Citation relations between scientific publications are represented as a simple undirected and unweighted graph by first discarding the directions of citations, any multiple citations and citations from a publication to itself. Publications neither citing nor cited by any other are also discarded. Let $n$ be the number of nodes $N$, $n=|N|$, and $m$ the number of links in such citation network. Denote $k$ to be the average node degree, i.e. the number of links incident to a node, $k=2m/n$, and LCC the largest connected component, i.e. the largest subset of mutually reachable nodes.

We analyze four citation networks representing publications in the fields of Scientometrics, Library \& Information Science and Physics, and also the entire science (see~\tblref{nets}). Publications and their citations were collected from the Web of Science bibliographic database produced by Thomson Reuters. More specifically, we used the in-house version of the Web of Science database of the Centre for Science and Technology Studies of Leiden University. This version of the Web of Science database is very similar to the one available online at \href{http://www.webofscience.com}{www.webofscience.com}. However, there are some differences, notably in the identification of citations between publications~\cite{OSV15}. Data collection was restricted to the Science Citation Index Expanded, the Social Sciences Citation Index and the Arts \& Humanities Citation Index, while only publications of the Web of Science document types `article' and `review' were included in the data collection.

\begin{table}[!h] \begin{adjustwidth}{-2.25in}{0in}
	\caption{{\bf Statistics of citation networks of scientific publications in Web of Science.} We consider three scientific fields and the entire Web of Science. See text for the definitions of the statistics and the details of the data collection procedure.}
	\begin{tabular}{llrrrrr} \hline
		{\bf Field} & {\bf Period} & {\bf \# Publications} & {\bf \# Nodes $n$} & {\bf \# Links $m$} &  {\bf Degree $k$} & {\bf \% LCC}  \\ \hline
		Scientometrics & 2009-2013 & $2$,$402$ & $1$,$998$ & $5$,$496$ & $5.50$ & $94.0\%$ \\ 
               	Library \& Infor.\ Sci.\ & 1996-2013 & $43$,$741$ & $32$,$628$ & $131$,$989$ & $8.09$ & $96.7\%$ \\ 
               	Physics & 2004-2013 & $1$,$314$,$458$ & $1$,$233$,$542$ & $9$,$838$,$008$ & $15.95$ & $98.5\%$ \\ 	
               	All Fields & 2004-2013 & $11$,$780$,$132$ & $11$,$063$,$916$ & $122$,$148$,$955$ & $22.08$ & $99.3\%$ \\ \hline
	\end{tabular}
	\label{tbl:nets}
\end{adjustwidth} \end{table}

The field of Scientometrics was delineated by selecting all publications in the following three journals: {\it Journal of Informetrics}, {\it Journal of the Association for Information Science and Technology} (including its precursor {\it Journal of the American Society for Information Science and Technology}), and {\it Scientometrics}. The field of Library \& Information Science was delineated by selecting all publications in the Web of Science journal subject category Information Science \& Library Science. Finally, the field of Physics was delineated by selecting all publications in the eight Physics journal subject categories in Web of Science as well as the subject category Astronomy \& Astrophysics.

\paragraph{Graph partitioning and community detection methods.} For a thorough empirical comparison, we select a large number of representative graph partitioning and community detection methods~\cite{For10, HDF14}, which we refer to as clustering methods in this paper. \tblref{methods} lists selected methods roughly divided into different classes. Due to the number of methods considered, detailed description is omitted here.

\begin{table}[!h] \begin{adjustwidth}{-2.25in}{0in}
	\caption{{\bf Graph partitioning and community detection methods.} We consider a large number of methods divided into different classes. See text for the details of methods implementation and parameters setting.}
	\begin{tabular}{lllc} \hline
		{\bf Class} & {\bf Method} & {\bf Description} & {\bf Ref.} \\ \hline
		Spectral analysis & Graclus & $k$-means clustering iteration & \cite{DGK07} \\ 
		& METIS & multi-level $k$-way partitioning & \cite{KK98a} \\ \hline
		Map equation~\cite{RB07} & Infomap & information flows compression & \cite{RB08} \\ 
		& Hiermap & hierarchical flows compression & \cite{RB11b} \\ \hline
		Modularity~\cite{NG04} & Louvain & greedy hierarchical optimization & \cite{BGLL08} \\ 
		& Mouvain & multi-level hierarchical optimization & \cite{RN11b} \\ 
		& SLM & smart local moving optimization & \cite{WV13} \\ \hline
		Label propagation & LPA & label propagation algorithm & \cite{RAK07} \\
		& BPA & balanced propagation algorithm & \cite{SB11d} \\
		& DPA & diffusion-propagation algorithm & \cite{SB11e} \\
		& HPA & hierarchical propagation algorithm & \cite{SB14a} \\
		& COPRA & community overlap propagation algorithm & \cite{Gre10} \\ \hline
		Statistical methods & OSLOM & order statistics local optimization method & \cite{LRRF11} \\ \hline
		Link clustering & Links & link similarity hierarchical clustering & \cite{ABL10} \\ \hline
		Graph models & BigClam & cluster affiliation matrix factorization & \cite{YL13} \\
		& CoDA & communities through directed affiliations & \cite{YML14} \\ \hline
		Ego-networks & DEMON & democratic estimate of modular organization & \cite{CRGP12} \\ \hline
		Random walks & Walktrap & random walks hierarchical clustering & \cite{PL05} \\ \hline
		Cliques & SCP & sequential clique percolation & \cite{KKKS08} \\
		& GCE & greedy clique expansion & \cite{LRMH10} \\\hline
	\end{tabular}
	\label{tbl:methods}
\end{adjustwidth} \end{table}

We use the source code provided by the authors of all methods in all cases except Mouvain and LPA, where we use our own implementations~\cite{WV13,SB11d}. We adopt default parameter settings of each particular algorithm. Graclus, METIS, BigClam and CoDA demand the number of clusters to be specified apriori. Thus, Graclus(S) and Graclus(L) denote the same method with the number of clusters set to $n/15$ and $n/50$, respectively, while Graclus refers to Graclus(S) on networks with $n<10^6$ and to Graclus(L) on larger networks (similarly for METIS, BigClam and CoDA). On the other hand, Links(S) and Links(L) denote the same method with Jaccard similarity threshold~\cite{ABL10} set to $0.1$ and $0.01$, respectively, whereas Links always refers to Links(S). Finally, some of the methods return overlapping clusters. For reasons of simplicity, each node in multiple clusters is assigned to the first cluster that appears in the output of the particular algorithm.

Certain otherwise prominent algorithms like Infomap can not be applied to very large networks in a time comparable with the fastest algorithms like Louvain and BPA. A straightforward solution is to first adopt some other method $\mathcal{M}$ to cut the network into smaller subgraphs and then independently apply Infomap to each of these. Let $C_i$ be some cluster of nodes in a network, $C_i\subseteq N$, and let $s_i$ be its size, $s_i=|C_i|$. Next, let $\mathcal{C}=\{C_i\}$ be the clustering of all the nodes in a network returned by the method $\mathcal{M}$, $\bigcup_i C_i=N$ and $C_i\cap C_j=\emptyset$, $i\neq j$. Then, for each cluster $C_i$ with $s_i>50$, Infomap is applied to the subgraph induced by the nodes in $C_i$, whereas the clustering of $C_i$ is accepted only when it improves the log-likelihood of $\mathcal{C}$ (see~\eqref{L}). Several such derived methods are considered. Gracmap and Metimap refer to methods that adopt spectral algorithms Graclus and METIS for the first method $\mathcal{M}$, respectively, where the number of clusters is set to $n/10^4$ for networks with $n<10^6$ and to $n/(5\cdot 10^4)$ otherwise. For comparison, we also include Louvmap and Labmap that adopt modularity optimization known as Louvain algorithm and label propagation algorithm LPA in the first step, respectively. Finally, the setting of the number of clusters in Graclus is limited to $2500$. Thus, for very large networks, we use Metilus that adopts METIS for $\mathcal{M}$ and Graclus afterwards. In total, we consider $30$ methods. These are the $20$ methods listed in \tblref{methods}, five variations with an alternative setting of the number of clusters and five derived methods as described above.

Let $\mathcal{C}=\{C_i\}$ be the clustering returned by some method $\mathcal{M}$. $\mathcal{C}$ often includes clusters $C_i$ that are too small or too large to be of any practical use, $s_i<s_{tiny}$ or $s_i>s_{giant}$. A straightforward solution is a two-step post-processing approach that first tries to further partition each of the giant clusters as above and then merges the tiny clusters with larger ones. We set $s_{tiny}=15$ and $s_{giant}=10^4$. First, for each cluster $C_i$ with $s_i>s_{giant}$, the same clustering method $\mathcal{M}$ is applied to the subgraph induced by the nodes in $C_i$ and the resulting clustering is accepted based on the log-likelihood of $\mathcal{C}$ as before. Note that, due to the resolution limit of community detection methods~\cite{FB07,TVN11}, most will further partition cluster $C_i$. Next, for each cluster $C_i$ with $s_i<s_{tiny}$, $C_i$ is merged with a neighboring cluster that most improves or least worsens the log-likelihood of $\mathcal{C}$. While the first post-processing step can be carried out simultaneously for each of the giant clusters, the tiny clusters in the second post-processing step have to be assessed in a random order.

\paragraph{Graph cuts and community structure statistics.} Let $\mathcal{C}$ be some clustering of network nodes as described above and let $A$ be the network adjacency matrix, $A_{ij}=A_{ji}\in\{0,1\}$ and $A_{ii}=0$. To measure the structure of clustering $\mathcal{C}$, we select different representative graph cuts and community structure statistics~\cite{YL12b}.
We measure the internal connectivity of clustering $\mathcal{C}$ as the average node internal degree $K$~\cite{RCCLP04},
\begin{equation} \label{eq:K} 
	K(\mathcal{C}) = \frac{1}{n}\sum_{ij} A_{ij} \delta(c_i,c_j),
\end{equation}
where $c_i$ is the cluster of node $i$ and $\delta$ is the Kronecker delta. The external connectivity of clustering $\mathcal{C}$ is measured as the average node external degree or expansion $E$~\cite{RCCLP04},
\begin{equation} \label{eq:E} 
	E(\mathcal{C}) = \frac{1}{n}\sum_{ij} A_{ij} (1-\delta(c_i,c_j)).
\end{equation}
By definition, $k=K+E$, whereas $K/k$ is the fraction of links covered by the clustering $\mathcal{C}$. Next, the Flake function $F$~\cite{FLG00} considers internal and external connectivity of clustering $\mathcal{C}$ and is defined as the fraction of nodes with larger external than internal degree,
\begin{equation} \label{eq:F} 
	F(\mathcal{C}) = \frac{\left|\left\{i: \sum_{j} A_{ij} \delta(c_i,c_j) < k_i/2 \right\}\right|}{n},
\end{equation}
where $k_i$ is the degree of node $i$. For reference with previous work, we also report the value of modularity function $Q$~\cite{NG04,New06b} that compares the internal connectivity of clustering $\mathcal{C}$ to the configuration model~\cite{NSW01}, i.e. a random graph with the same degree sequence,
\begin{equation} \label{eq:Q} 
	Q(\mathcal{C}) = \frac{1}{2m}\sum_{ij} \left(A_{ij} - \frac{k_ik_j}{2m}\right)\delta(c_i,c_j).
\end{equation}

Finally, we report the posterior probability of clustering $\mathcal{C}$ or the likelihood of $\mathcal{C}$ given the network observed~\cite{Pei12c}. Assume that links in a network formed solely based on nodes' cluster membership and let $\theta_i$ be a linking probability associated with cluster $C_i$. Then $m_i$ links observed between the nodes in cluster $C_i$ would form with probability $\theta_i^{m_i}$ and the remaining $M_i-m_i$ possible links would not form with probability $(1-\theta_i)^{M_i-m_i}$, $M_i=s_i(s_i-1)/2$. Let $\tilde{\theta}$ be a linking probability representing the connectivity between the clusters. Then $\widetilde{m}$ links observed between the nodes in different clusters would form with probability $\tilde{\theta}^{\widetilde{m}}$, $\widetilde{m}=m-\sum_im_i$, and the remaining $\widetilde{M}-\widetilde{m}$ possible links would not form with probability $(1-\tilde{\theta})^{\widetilde{M}-\widetilde{m}}$, $\widetilde{M}=n(n-1)/2-\sum_iM_i$. Thus, the probability that the network formed according to $\mathcal{C}$ or the likelihood of $\mathcal{C}$ is defined as
\begin{equation} \label{eq:L} 
	L(\mathcal{C})=\tilde{\theta}^{\widetilde{m}}(1-\tilde{\theta})^{\widetilde{M}-\widetilde{m}}\prod_i\theta_i^{m_i}(1-\theta_i)^{M_i-m_i},
\end{equation}
where $\theta_i=m_i/M_i$ and $\tilde{\theta}=\widetilde{m}/\widetilde{M}$ are the maximum likelihood estimators~\cite{CB90}. For reasons of numerical stability, we report the log-likelihood of $\mathcal{C}$ as $\log{L}(\mathcal{C})$.

Denote $C$ to be a random variable corresponding to clustering $\mathcal{C}$, $\mathrm{P}(C=C_i)=s_i/n$. The distance between two clusterings $\mathcal{C}$ and $\mathcal{D}$ is measured using the variation of information $V$~\cite{Mei07} defined as
\begin{equation} \label{eq:V} 
	V(\mathcal{C},\mathcal{D}) = H(C|D) + H(D|C),
\end{equation}
where $H(C|D)$ and $H(D|C)$ are conditional entropies. Since $V\in[0,\log{n}]$, we report the normalized variation of information $V/\log{n}$~\cite{KLN08}.

Clustering robustness plots $R(\mathcal{M},\alpha)$~\cite{KLN08} estimate the robustness of clustering $\mathcal{C}$ or the respective clustering method $\mathcal{M}$ under random perturbations of network links. $R$ is defined as the distance between $\mathcal{C}$ and $\mathcal{C}_{\alpha}$, 
\begin{equation} \label{eq:R} 
	R(\mathcal{M},\alpha) = V (\mathcal{C}, \alpha) = V(\mathcal{C},\mathcal{C}_{\alpha}),
\end{equation}
where $\mathcal{C}_{\alpha}$ is obtained by $\mathcal{M}$ after randomly rewiring $\alpha$ links in the network. 

% All values, plots and diagrams reported in~\resref are averages over $100$ realizations for Scientometrics, $10$ realizations for Library \& Information Science, two realizations for Physics and a single realization for All Fields.

\paragraph{Bibliometric clustering criteria.} Let $\mathcal{C}$ be some clustering of network nodes as described above. To measure the utility of clustering $\mathcal{C}$, we select different bibliometric clustering criteria. We report the average cluster size $S$ and the fraction of covered links $K/k$ already introduced above. Next, we define the orders of magnitude covered by cluster sizes $O$ as
\begin{equation} \label{eq:O} 
	O(\mathcal{C}) = \log_{10}{\frac{s_L}{s_S}},
\end{equation}
where $s_L$ is the size of the largest cluster and $s_S$ is the size of the smallest. Note that twice the value of $s_S$, which is negligible, has the same effect on $O$ as twice the value of $s_L$, which is substantial. We thus report 5-percentile effective orders $O_5$ defined as
\begin{equation} \label{eq:O5} 
	O_5(\mathcal{C}) = \log_{10}{\frac{s_L}{s_5}},
\end{equation}
where $s_5$ is the size of the smallest remaining cluster after removing the $5\%$ smallest clusters. To measure the diameter of clusters in $\mathcal{C}$, we compute the $90$-percentile effective cluster diameter $D_{90}$~\cite{LKF07}, i.e. the average number of hops to reach $90\%$ of all the nodes within a cluster. The value of $D_{90}$ is estimated from $1000$ randomly selected seed nodes. Finally, the robustness of clustering $\mathcal{C}$~\cite{KLN08} or equivalently the uncertainty $U$ of the respective clustering method $\mathcal{M}$ is defined as the distance between the clusterings $\mathcal{C}_1$ and $\mathcal{C}_2$ obtained by two consecutive realizations of $\mathcal{M}$ (see~\eqref{V}),
\begin{equation} \label{eq:U} 
	U(\mathcal{M}) = V(\mathcal{C}_1,\mathcal{C}_2).
\end{equation}

All values, plots and diagrams reported in~\resref are averages over $100$ realizations for Scientometrics, $10$ realizations for Library \& Information Science, two realizations for Physics and a single realization for All Fields.

% % % % % % % % % % % % % % % % % % % % % % % % % % 
%
%					RESULTS
%
% % % % % % % % % % % % % % % % % % % % % % % % % %

\section*{Results}

We start by directly comparing the clusterings obtained by all $30$ clustering methods described in~\metref to derive a manageable set of representatives. Next, we analyze structural and bibliometric statistics of the clusterings obtained by representative methods, and perform an expert-based assessment of the clusterings. Last, we analyze also the large-scale behavior of the most prominent methods.

\begin{figure}[!h] 
	\includegraphics[width=1.0\textwidth]{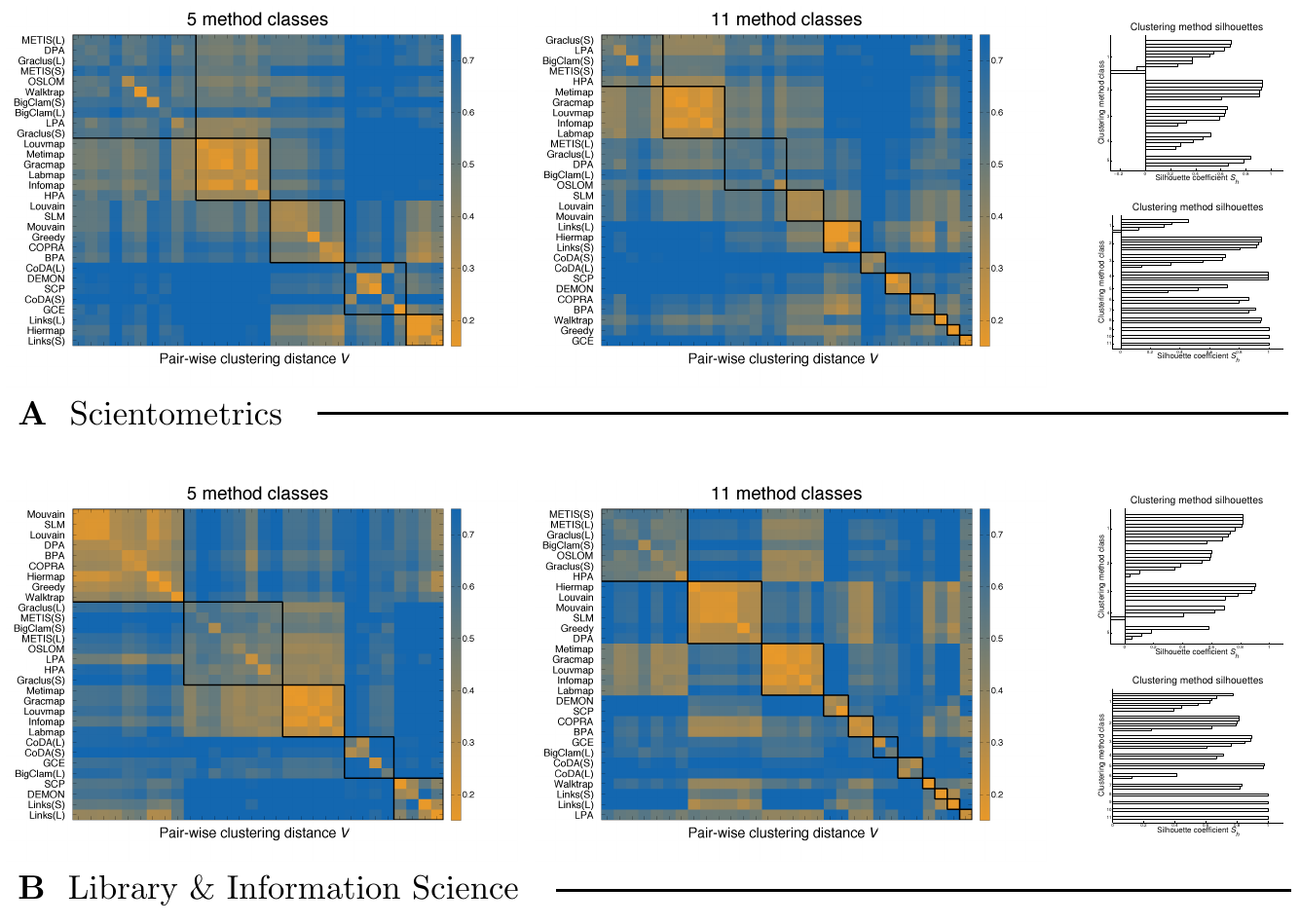} \\
%	\includegraphics[width=0.4\textwidth]{scientometrics_2009_2013_5}
%	\includegraphics[width=0.4\textwidth]{scientometrics_2009_2013_11} ~
%	\begin{minipage}[b][0.2875\textwidth][s]{0.1475\textwidth} \centering
%		\includegraphics[width=\textwidth]{scientometrics_2009_2013_5_silhouette} \\[3pt]
%		\includegraphics[width=\textwidth]{scientometrics_2009_2013_11_silhouette} 
%	\end{minipage}
%	\tikz{\node{{\bf A}~~Scientometrics}; \draw[thick] (1.75,0) -- (11.875,0);} \\~\\
%	\includegraphics[width=0.4\textwidth]{lis_1996_2013_5}
%	\includegraphics[width=0.4\textwidth]{lis_1996_2013_11} ~
%	\begin{minipage}[b][0.2875\textwidth][s]{0.1475\textwidth} \centering
%		\includegraphics[width=\textwidth]{lis_1996_2013_5_silhouette} \\[3pt]
%		\includegraphics[width=\textwidth]{lis_1996_2013_11_silhouette}
%	\end{minipage}
%	\tikz{\node{{\bf B}~~Library \& Information Science}; \draw[thick] (3,0) -- (10.675,0);} \\
	\caption{{\bf Pair-wise distances between the clusterings obtained by the considered methods.} Panel~{\bf A} shows the heatmaps of clustering distances for the Scientometrics citation network, where the methods are clustered into $5$ and $11$ classes (left- and right-hand side, respectively). Note that this merely implies the ordering of the rows/columns. Insets on the right show the method silhouette coefficients. Panel~{\bf B} shows the same for the Library \& Information Science citation network. See~\metref for the definition of the clustering distance and text for the details of the method clustering procedure.}
	\label{fig:classes}
\end{figure}

\paragraph{Pair-wise clustering comparison.} \figref{classes} shows heatmaps of the pair-wise distances between the clusterings returned by the considered methods (see~\eqref{V}). The methods are applied to two citation networks representing the fields of Scientometrics and Library \& Information Science (see~\tblref{nets}). To gain insight into different classes of methods, we apply the $k$-means data clustering algorithm~\cite{Mac67} to the rows/columns of the heatmaps  with the number of classes set to $5$ and $11$ (left- and right-hand side of~\figref{classes}, respectively). The classes of methods are shown in the order of decreasing size and the methods within each class are listed in the order of decreasing silhouette coefficients $S_h$~\cite{Rou87}. $S_h(\mathcal{M})$ of some method $\mathcal{M}$ is defined as a normalized difference between the lowest average inter-class dissimilarity and the average intra-class dissimilarity, for which we adopt the standard cosine similarity.

We observe compact classes of methods, most notably pronounced for the larger network (see right-hand side of~\figref{classes}, panel~{\bf B}). Namely, the largest three classes represent spectral and statistical methods (e.g. Graclus, METIS and OSLOM), modularity optimization (e.g. Louvain and SLM) and map equation algorithms (e.g. Gracmap, Metimap and Infomap). Other smaller classes correspond to label propagation algorithms (e.g. LPA, BPA and COPRA), random walks (e.g. Walktrap), link clustering (i.e. Links), methods based on cliques (i.e. GCE and SCP) and other methods. Thus, despite the large number of methods considered, these can be divided into only a handful of truly different classes, but the differences between the classes can be rather substantial. In the following we limit the analysis to the $15$ class representatives explicitly stated above, although the actual subset of methods considered depends on the size of the network analyzed.

\paragraph{Structural clustering analysis.} Past literature often reported a power-law form $s^{-\gamma}$ of the cluster size distribution $\mathrm{P}(s)$~\cite{New04b,CNM04}, to the extent that $s^{-\gamma}$ is also incorporated into the standard network benchmarks for testing clustering methods~\cite{LFR08,LF09a}. Nevertheless, this may be merely an artifact of the power-law degree distribution $\mathrm{P}(k)\sim k^{-\gamma}$ observed in real-world networks~\cite{KN11b}, while recent work on principled clustering methods sheds further doubts on the power-law form of $\mathrm{P}(s)$~\cite{Pei15}.

\begin{figure}[!b] 
	\includegraphics[width=1.0\textwidth]{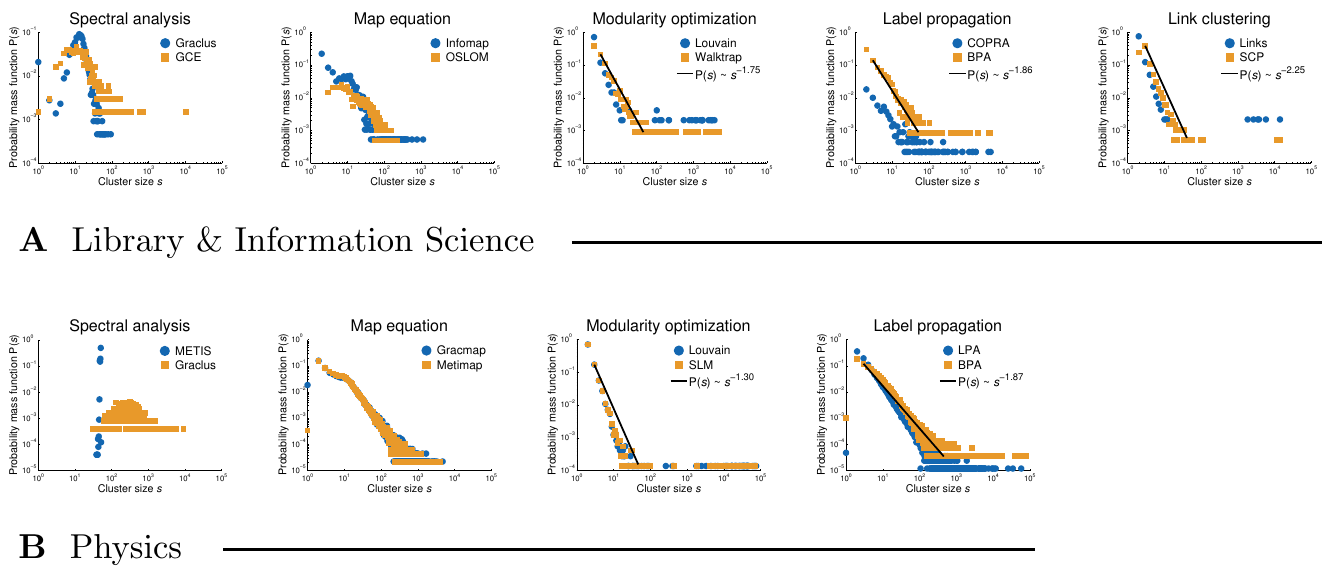} \\
%	\includegraphics[width=0.17\textwidth]{lis_1996_2013_size_specter} ~~~~
%	\includegraphics[width=0.17\textwidth]{lis_1996_2013_size_map} ~~~~
%	\includegraphics[width=0.17\textwidth]{lis_1996_2013_size_Q} ~~~~
%	\includegraphics[width=0.17\textwidth]{lis_1996_2013_size_label} ~~~~
%	\includegraphics[width=0.17\textwidth]{lis_1996_2013_size_link} \\[7pt]
%	\tikz{\node{{\bf A}~~Library \& Information Science}; \draw[thick] (3,0) -- (10.625,0);} \\~\\
%	\includegraphics[width=0.17\textwidth]{physics_2004_2013_size_specter} ~~
%	\includegraphics[width=0.17\textwidth]{physics_2004_2013_size_map} ~~
%	\includegraphics[width=0.17\textwidth]{physics_2004_2013_size_Q} ~~
%	\includegraphics[width=0.17\textwidth]{physics_2004_2013_size_label} \\[7pt]
%	\tikz{\node{{\bf B}~~Physics}; \draw[thick] (1.25,0) -- (9.5,0);} \\ 
	\caption{{\bf Size distributions of the clusterings obtained by representative methods.} Panels~{\bf A} and~{\bf B} show cluster size distributions $\mathrm{P}(s)$ for the Library \& Information Science and Physics citation networks, respectively. Wherever plausible, the power-laws $s^{-\gamma}$ are fitted to the tails of the distributions by maximum likelihood estimation, $\gamma = 1+n\left(\sum_i \log{s_i/s_{min}}\right)$ for $s_{min}>1$.}
	\label{fig:sizes} 
\end{figure} 

\figref{sizes} shows the distributions $\mathrm{P}(s)$ of the clusterings returned by representative methods applied to the Library \& Information Science and Physics citation networks (see~\tblref{nets}). The methods are paired according to a similar shape of $\mathrm{P}(s)$, where each pair is named by its most ``famous'' representative. Statistical methods are thus reported under map equation, while methods based on cliques appear under spectral analysis and link clustering. Notice that the validity of the power-law claim $\mathrm{P}(s)\sim s^{-\gamma}$ clearly depends on the particular method considered. For instance, there is evidently a peek in the distributions of spectral methods with a lack of heavy tail (see left-hand side of~\figref{sizes}, panel~{\bf A}). Furthermore, in the case of map equation and statistical methods, the power-law form $s^{-\gamma}$ is violated for small and moderate $s$. On the other hand, the distributions for modularity optimization, label propagation and link clustering seem to follow the power-law scaling over several orders (see right-hand side of~\figref{sizes}, panel~{\bf A}) with the power-law exponent $\gamma$ increasing from left to right. In the extreme case, link clustering produces a few very large clusters covering most of the nodes in the network, while the size distribution of the remaining ones follows a power-law. The observed differences between the clustering methods are even more striking on a larger network (see~\figref{sizes}, panel~{\bf B}).

\tblref{stats} shows structural statistics of the clusterings obtained by representative methods applied to the Library \& Information Science citation network. Most methods return a little less than $2000$ clusters with some notable exceptions. Modularity optimization method Louvain, and also the methods based on dynamical processes (e.g. Walktrap and BPA), return a much smaller number clusters. % which can be attributed to their resolution limit~\cite{FB07,TVN11}. 
On the other hand, link clustering and some other methods (e.g. COPRA) return a much larger number of clusters.

\begin{table}[!h] \begin{adjustwidth}{-2.25in}{0in}
	\caption{{\bf Structural statistics of the clusterings obtained by representative methods.} The methods are applied to the Library \& Information Science citation network. See~\metref for the definitions of the statistics and text for the interpretation.}
	\begin{tabular}{lrrrrrr} \hline
		{\bf Method} & {\bf \# Clusters}& {\bf Degree $K$} & {\bf Expansion $E$} & {\bf Flake $F$} & {\bf Modularity $Q$} &  {\bf Likelihood $\log L$}  \\ \hline
		Louvain & $488.2$ & $6.81$ & $1.28$ & $3.3\%$ & $0.734$ & $-978498.8$ \\ 
               	GCE & $682.0$ & $4.06$ & $4.03$ & $28.9\%$ & $0.431$ & $-997346.0$ \\ 
               	BPA & $1001.9$ & $7.00$ & $1.09$ & $3.0\%$ & $0.664$ & $-975063.7$ \\ 
               	Walktrap & $1127.0$ & $6.47$ & $1.62$ & $7.0\%$ & $0.686$ & $-968783.9$ \\ 		
               	Infomap & $1871.2$ & $5.00$ & $3.09$ & $19.3\%$ & $0.602$ & $-836963.9$ \\ 
               	OSLOM & $1914.2$ & $3.79$ & $4.30$ & $36.9\%$ & $0.453$ & $-932170.7$ \\ 
               	SCP & $1969.0$ & $4.92$ & $3.17$ & $37.2\%$ & $0.217$ & $-1103053.0$ \\ 
               	Graclus & $2175.0$ & $2.36$ & $5.73$ & $52.4\%$ & $0.290$ & $-1003511.5$ \\ 
               	% Links & $2590.4$ & $5.91$ & $2.18$ & $24.7\%$ & $0.125$ & $-1171165.4$ \\ 
		Links & $2933.1$ & $6.39$ & $1.70$ & $20.0\%$ & $0.093$ & $-1173310.5$ \\
               	COPRA & $3825.5$ & $6.83$ & $1.26$ & $15.1\%$ & $0.645$ & $-993909.5$ \\ \hline
	\end{tabular}
	\label{tbl:stats}
\end{adjustwidth} \end{table}

\tblref{stats} further shows the average internal degree of the nodes in the clusters $K$ and the average external degree or expansion $E$ (see~\eqsref{K}{E}). % Although all except the spectral method Graclus achieve $K\gg E$
Although most methods achieve $K\gg E$, there are some important differences between the methods. The Flake function $F$ measures the fraction of nodes with larger external than internal cluster degree (see~\eqref{F}). Notice that the values of $F$ reflect the differences in the cluster size distributions $\mathrm{P}(s)$ observed in~\figref{sizes}. Modularity optimization and other methods that return clusterings with a power-law distribution $\mathrm{P}(s)\sim s^{-\gamma}$ can, due to a number of very large clusters, effectively cover many of the links in the network, giving low $F$ (e.g. Louvain, Walktrap and BPA). On the contrary, spectral methods with a rather homogeneous distribution $\mathrm{P}(s)$ must inevitably cut a large number of links between the clusters, thus giving very high $F$ (e.g. Graclus). As in~\figref{sizes}, the middle ground between these two regimes is represented by map equation and statistical methods (e.g. Infomap and OSLOM).

Mainly for reference with previous work, \tblref{stats} shows the values of modularity $Q$ (see~\eqref{Q}). Expectedly, the modularity optimization method Louvain gives the highest $Q$. \tblref{stats} also reports the log-likelihood $\log{L}$ of the clusterings given the network observed (see~\eqref{L}). The most likely clustering is obtained by Infomap, yet it should be stressed that the map equation is actually a likelihood criterion.

\begin{figure}[!t]
	\includegraphics[width=1.0\textwidth]{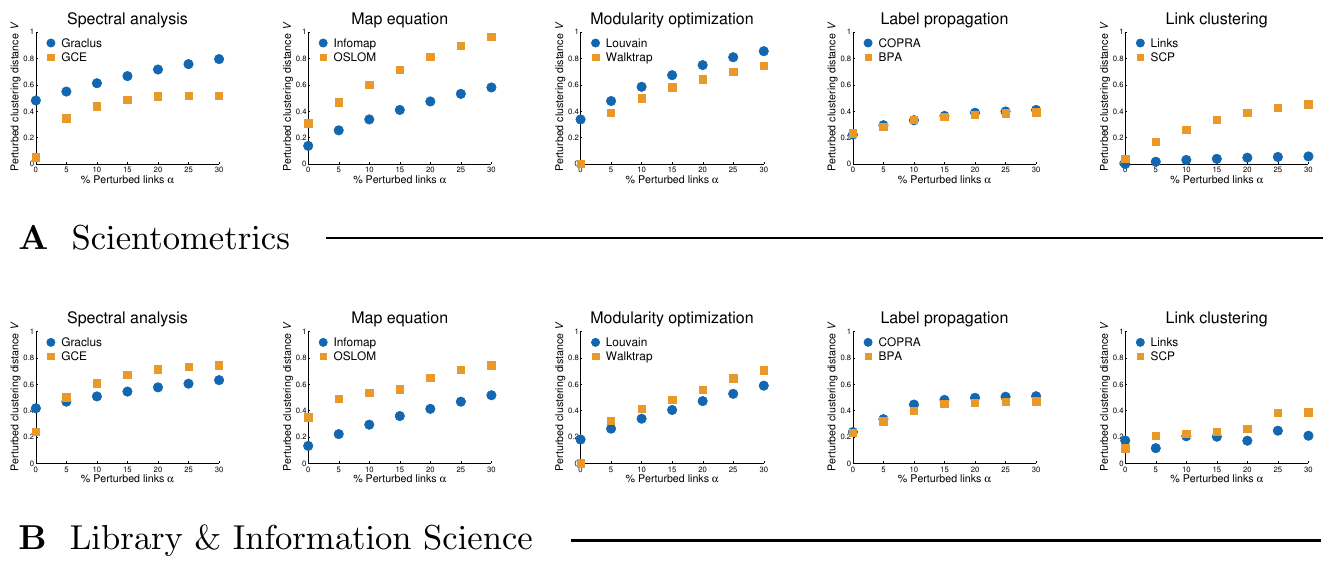} \\
%	\includegraphics[width=0.17\textwidth]{scientometrics_2009_2013_robust_specter} ~~~~
%	\includegraphics[width=0.17\textwidth]{scientometrics_2009_2013_robust_map} ~~~~
%	\includegraphics[width=0.17\textwidth]{scientometrics_2009_2013_robust_Q} ~~~~
%	\includegraphics[width=0.17\textwidth]{scientometrics_2009_2013_robust_label} ~~~~
%	\includegraphics[width=0.17\textwidth]{scientometrics_2009_2013_robust_link} \\[7pt]
%	\tikz{\node{{\bf A}~~Scientometrics}; \draw[thick] (1.75,0) -- (11.875,0);} \\~\\
%	\includegraphics[width=0.17\textwidth]{lis_1996_2013_robust_specter} ~~~~
%	\includegraphics[width=0.17\textwidth]{lis_1996_2013_robust_map} ~~~~
%	\includegraphics[width=0.17\textwidth]{lis_1996_2013_robust_Q} ~~~~
%	\includegraphics[width=0.17\textwidth]{lis_1996_2013_robust_label} ~~~~
%	\includegraphics[width=0.17\textwidth]{lis_1996_2013_robust_link} \\[7pt]
%	\tikz{\node{{\bf B}~~Library \& Information Science}; \draw[thick] (3,0) -- (10.625,0);} \\
	\caption{{\bf Robustness of the clusterings obtained by representative methods.} Panels~{\bf A} and~{\bf B} show clustering robustness plots $V(\alpha)$ for the Scientometrics and Library \& Information Science citation networks, respectively. These show the distances between the clusterings obtained after randomly rewiring $\alpha$ links. See~\metref for the definitions of clustering distance and robustness.}
	\label{fig:robs}
\end{figure}

\figref{robs} shows the robustness plots $V(\alpha)$ of the clusterings returned by representative methods for the Scientometrics and Library \& Information Science citation networks (see~\eqref{R}). The plots measure the distances between the clusterings obtained by the same method after randomly rewiring $\alpha$ links in the network. Although initially introduced as a measure of network community structure~\cite{KLN08}, we here adopt the same approach to measure the robustness of different clusterings.

The methods in~\figref{robs} are paired as in~\figref{sizes}. Since many of them are nondeterministic, most of the plots do not start in the origin. The clusterings obtained by spectral and statistical methods (e.g. Graclus and OSLOM) prove to be the least robust with high values of $V$ even for small $\alpha$ (see left-hand side of~\figref{robs}). Map equation algorithm Infomap, and modularity optimization on the larger network (see middle of~\figref{robs}, panel~{\bf B}), seem to give stable clusterings with gradually increasing $V$ over all $\alpha$. Label propagation methods and link clustering appear very robust at first sight with surprisingly low $V$ even for very large $\alpha$ (see right-hand side of~\figref{robs}). For instance, the clustering returned by Links stays almost unchanged even after rewiring $30\%$ of the links in the network. Nevertheless, this is a consequence of the existence of a few very large clusters that occupy the majority of the nodes in the network (see~\figsref{sizes}{degs}) and change very little compared to the clusterings returned by other methods.

\paragraph{Bibliometric clustering analysis.} The above structural analysis of the clusterings of citation networks would most likely be of interest to network scientists, but might provide limited value to the bibliometric community. In the following, we therefore analyze the clusterings also from an alternative perspective.

\tblref{bibs} shows bibliometric statistics of the clusterings obtained by representative methods applied to the Library \& Information Science citation network. The average cluster sizes $S$ can be interpreted as the number of clusters in~\tblref{stats}. For most methods, $S\approx 15$. % Due to the resolution limit, 
Modularity optimization method Louvain gives almost five times larger clusters on average, while link clustering and some other methods (e.g. COPRA) return much smaller clusters with $S\approx 10$. \tblref{bibs} further shows $5$-percentile effective orders $O_5$ that measure the orders of magnitude covered by cluster sizes $s$ (see~\eqref{O5}). For many practical applications, the clusters ideally should span no more than a single order of magnitude giving $O_5\approx 1$. This turns out to be an illusive goal as $O_5\gg 1$ for all methods except the spectral ones (e.g. Graclus), which one can observe also in~\figref{sizes}. Next, the $90$-percentile effective diameter $D_{90}$ measures the average number of hops to reach most of the nodes in a cluster (see~\metref). Most methods return clusterings with small $D_{90}$ consistent with the small-world network structure~\cite{WS98}. On the other hand, $D_{90}>10$ for methods based on cliques (i.e. GCE and SCP) and link clustering, indicating the existence of some very large clusters, which is rather inconvenient in practice.

\begin{table}[!h] \begin{adjustwidth}{-2.25in}{0in}
	\caption{{\bf Bibliometric statistics of the clusterings obtained by representative methods.} The methods are applied to the Library \& Information Science citation network. See~\metref for the definitions of the statistics and text for the interpretation.}
	\begin{tabular}{lrrrrrr} \hline
		{\bf Method} & {\bf Size $S$} & {\bf Orders $O_{5}$} & {\bf Diameter $D_{90}$} & {\bf Coverage $K/k$} & {\bf Uncertainty $U$} &  {\bf Complexity $T$}  \\ \hline
		Louvain & $66.7$ & $3.33$ & $9.13$ & $84.5\%$ & $0.194$ & $0.6$ sec \\ 
               	GCE & $47.8$ & $3.32$ & $11.99$ & $50.1\%$ & $0.241$ & $26.5$ sec \\ 
               	BPA & $32.0$ & $3.61$ & $7.28$ & $86.2\%$ & $0.213$ & $3.3$ sec \\ 
               	Walktrap & $29.0$ & $3.39$ & $7.80$ & $79.9\%$ & $0.000$ & $34.9$ sec \\ 
		% Links & $22.3$ & $4.34$ & $11.54$ & $77.3\%$ & $0.170$ & $9.7$ sec \\ 
               	Infomap & $17.3$ & $2.68$ & $4.32$ & $61.5\%$ & $0.133$ & $9.6$ sec \\ 
		SCP & $16.6$ & $4.15$ & $23.12$ & $60.8\%$ & $0.021$ & $1.4$ sec \\ 
               	OSLOM & $16.0$ & $2.61$ & $4.82$ & $45.9\%$ & $0.364$ & $94.9$ sec \\ 
               	Graclus & $15.0$ & $1.13$ & $3.38$ & $29.2\%$ & $0.417$ & $6.4$ sec \\ 
		Links & $10.1$ & $4.31$ & $11.09$ & $78.0\%$ & $0.048$ & $10.0$ sec \\
               	COPRA & $8.8$ & $3.97$ & $6.91$ & $84.9\%$ & $0.217$ & $27.0$ sec \\ \hline
	\end{tabular}
	\label{tbl:bibs}
\end{adjustwidth} \end{table}

\tblref{bibs} also shows the fractions of the links covered by different clusterings $K/k$ (see~\metref). Notice substantial diversity between the methods, which can again be interpreted in terms of different cluster size distributions $\mathrm{P}(s)$ (see~\figref{sizes}). The methods that return clusterings with a power law $\mathrm{P}(s)\sim s^{-\gamma}$, namely modularity optimization (e.g. Louvain), link clustering and methods based on dynamical processes (e.g. Walktrap, COPRA and BPA), can effectively cover over $80\%$ of the links in the network. However, spectral and statistical methods (e.g. Graclus and OSLOM) that are characterized by a rather homogeneous $\mathrm{P}(s)$ give $K/k$ as low as $30\%$. The middle ground is again represented by the map equation algorithm Infomap with $K/k$ around~$60\%$.

The uncertainty $U$ measures the stability of a method or equivalently the distance between the clusterings obtained by two consecutive realizations of the same method (see~\eqref{U}). Note that $U=V(0)$ in~\figref{robs}. \tblref{bibs} shows the uncertainties of representative clustering methods. Spectral and statistical methods (e.g. Graclus and OSLOM) are substantially less stable than the rest with $U\approx 0.4$. Due to the existence of a few very large clusters already discussed above, link clustering and some other methods (i.e. Walktrap and SCP) appear very robust with $U\approx 0$. For the rest, $U\approx 0.2$.

The method complexity $T$ in~\tblref{bibs} is measured as the execution time on a $2.3$ GHz Intel Core i7 processor with a sufficient amount of memory. The fastest methods are those based on modularity optimization (i.e. Louvain), label propagation (e.g. BPA) and also spectral analysis (e.g. Graclus). Notice that the map equation algorithm Infomap takes only about ten seconds on the Library \& Information Science citation network. Although this does not seem much, the network is relatively small. In fact, the algorithm takes almost three hours on the Physics citation network (results not shown) and would probably take several days to cluster the All Fields citation network (see~\tblref{nets}).

% Clustering degeneracy diagrams $D(\mathcal{C})$ display non-degenerate or effective range of clustering $\mathcal{C}$ defined as the pair in~\eqref{D}.
% \begin{equation} \label{eq:D} 
% 	D(\mathcal{C}) = \left(\sum_{s_i<s_{tiny}} s_i,n-s_L\right)
% \end{equation}
% \figref{degs} shows the degeneracy diagrams $D$ of the clusterings returned by representative methods on Library \& Information Science and Physics citation networks (see~\eqref{D}). The plots display the non-degenerate or effective ranges of the clusterings comprising the fraction of nodes not contained in tiny clusters with $s<s_{tiny}$, $s_{tiny}=15$, and in the largest or giant cluster. In the best-case scenario, the ranges in~\figref{degs} would span from top to bottom. Any deviation from the top or bottom signifies the existence of at least one very large cluster or many tiny clusters, respectively.

\begin{figure}[!t] 
	\includegraphics[width=1.0\textwidth]{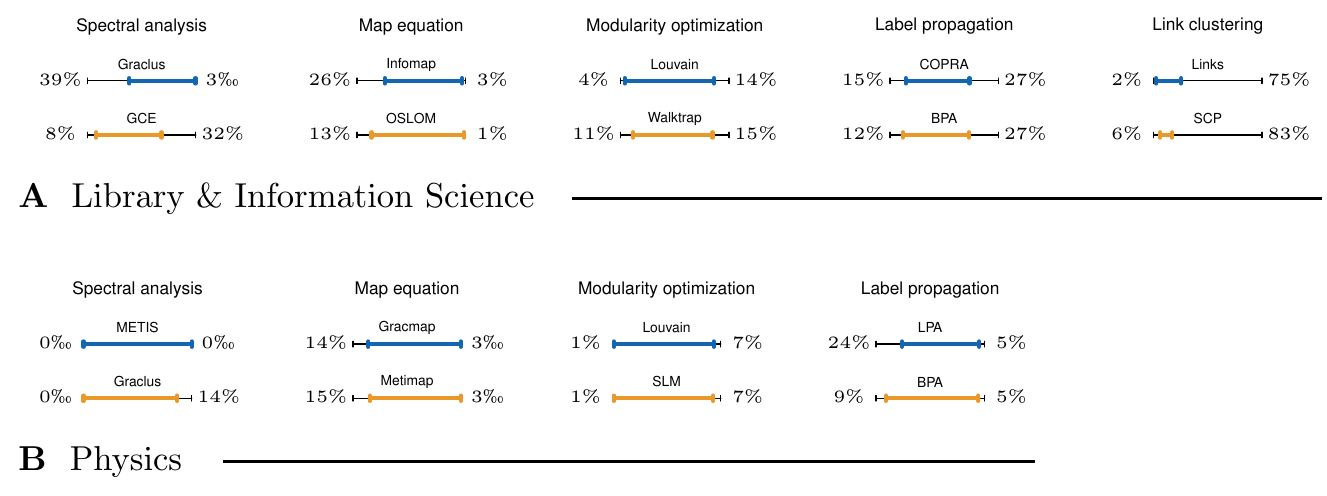} \\
	\caption{{\bf Degeneracy of the clusterings obtained by representative methods.} Panels~{\bf A} and~{\bf B} show clustering degeneracy diagrams $D$ for the Library \& Information Science and Physics citation networks, respectively. These display the non-degenerate ranges of the clusterings, while the percentages show the fraction of nodes in tiny clusters $\sum_{s_i<s_{tiny}} s_i/n$ and in the largest cluster $s_L/n$ (left- and right-hand side, respectively). See text for the definition of clustering degeneracy.}
	\label{fig:degs}
\end{figure}

\figref{degs} shows the degeneracy diagrams $D$ of the clusterings returned by representative methods on the Library \& Information Science and Physics citation networks. These display the non-degenerate or effective ranges of the clusterings that span the fraction of nodes not covered by tiny clusters with $s<s_{tiny}$, $s_{tiny}=15$, or the largest or giant cluster. Hence, the degeneracy diagram $D$ is defined as a range $(\sum_{s_i<s_{tiny}} s_i/n,1-s_L/n)$, where $s_L$ is the size of the largest cluster. In the best-case scenario, the ranges in~\figref{degs} would span from left to right. Any deviation from right or left signifies the existence of at least one very large cluster or many tiny clusters, respectively. 

The methods in~\figref{degs} are paired as in~\figref{sizes}. The map equation algorithm Infomap and spectral and statistical methods (e.g. Graclus and OSLOM) return clusterings without a giant cluster spanning a large fraction of the nodes (see left-hand side of~\figref{degs}, panel~{\bf A}). However, these can include many tiny clusters. On the other hand, modularity optimization and label propagation methods (e.g. Louvain and BPA) return clusterings with at least one very large cluster (see right-hand side of~\figref{degs}, panel~{\bf A}). Even more, in the case of link clustering and some other methods (e.g. SCP), the giant cluster contains almost all the nodes in the network. Although the existence of a giant cluster and tiny clusters is not clearly visible in the case of a larger network (see~\figref{degs}, panel~{\bf B}), we stress that even a slight deviation from right or left is already substantial.

\paragraph{Expert-based clustering assessment.} An expert-based assessment was performed on the clusterings obtained by representative methods on the Library \& Information Science citation network. Within this network, the assessment focused on clusters covering topics or research areas in the field of scientometrics. Scientometrics can be seen as a subfield of the broader field of library and information science. The assessment was performed jointly by the second and the third author (NJvE and LW), who both have an extensive expertise in the field of scientometrics. A detailed investigation and comparison of the different clusterings was done with the help of the CitNetExplorer software tool for visualizing and analyzing citation networks of publications~\cite{VW14}.

We start by comparing the obtained clusterings based on the resolution they provide. A clustering consisting of a small number of clusters, with each cluster including a relatively large number of publications, has a low resolution. On the other hand, a clustering consisting of a large number of clusters, each including only a small number of publications, has a high resolution.

There are a number of clusterings for which we consider the resolution to be too high. This is the case for spectral methods Graclus(S), Graclus(L), METIS(S) and METIS(L). In these clusterings, topics that we would expect to be represented by a single cluster were instead represented by multiple clusters, each covering a subset of the publications dealing with a topic. For instance, the clustering returned by Graclus(L) includes four clusters that all cover part of the literature on the topic of the h-index, a very prominent topic in the field of scientometrics. Of these four clusters, there is one that clearly has its own focus. This cluster includes publications studying the mathematical properties of the h-index. Having a separate cluster for these publications is probably defensible. However, the other three clusters all seem to cover very similar publications, and therefore we see no justification for the fact that these publications are distributed over three clusters rather than all being assigned to the same cluster.

Other clusterings have a resolution that is too low for a meaningful analysis of the scientometric literature. The clusterings for which this is the case are obtained by BPA and Walktrap. One of the clusters created by BPA for instance consists of $3$,$808$ publications and essentially covers the entire scientometric literature. This cluster seems to properly delineate the scientometric literature from the rest of the library and information science literature. Hence, if one’s purpose is to identify subfields within the field of library and information science, then BPA may provide good results. However, in our case, we are interested in identifying topics rather than entire subfields, and for this purpose the results provided by BPA are not helpful.

The clusterings with a resolution that matches reasonably well with the idea of identifying topics within the subfield of scientometrics are obtained by the statistical method OSLOM and the map equation algorithms Infomap and Metimap. In addition to the clustering methods presented in~\metref, we here consider also a variant of the Louvain modularity optimization method with a resolution parameter~\cite{RB06a} that one can tune to customize the clustering resolution~\cite{WV13}. Setting the resolution parameter to $10$ gives the most suitable resolution here, which we denote Louvain($10$). We next analyze OSLOM, Infomap, Metimap and Louvain($10$) in more detail.

The clustering obtained by OSLOM has a relatively high resolution. It includes only three clusters with more than $100$ scientometric publications, which means that most scientometric publications are assigned to small clusters. As a consequence, some topics that we would expect to be represented by a single cluster are in fact distributed over multiple clusters. Important examples are the topic of webometrics and the topic of patents. These topics are each distributed over two clusters of approximately equal size, which we consider an unsatisfactory result. A more general problem of OSLOM is that we observe a relatively large number of publications that are assigned to a cluster where they do not seem to belong. For instance, there is a cluster covering the topic of the analysis and visualization of bibliometric networks, but this cluster includes a significant number of publications dealing with other topics, such as the topic of indicators for citation analysis.

Louvain($10$) clustering is characterized by a somewhat unusual cluster size distribution. Compared with other clusterings, it includes a relatively large number of clusters with more than $100$ publications and a relatively small number of clusters with a number of publications between $10$ and $100$. As a consequence, there are a number of larger scientometric clusters for which there is no similar cluster in other clusterings, for instance obtained by Metimap or Infomap. A detailed examination of these clusters indicates that they do not cover easily recognizable topics. Publications included in these clusters usually do have something in common. For instance, there are clusters in which many publications relate to a specific country or a specific geographical region, such as China or Africa. However, our overall impression is that the clusters are of a somewhat heterogeneous nature and that it would have been better if the publications in the clusters had been distributed over a number of smaller clusters. The presence of these heterogeneous clusters is a significant weakness of Louvain($10$).

\begin{figure}[!h] 
	\includegraphics[width=1.0\textwidth]{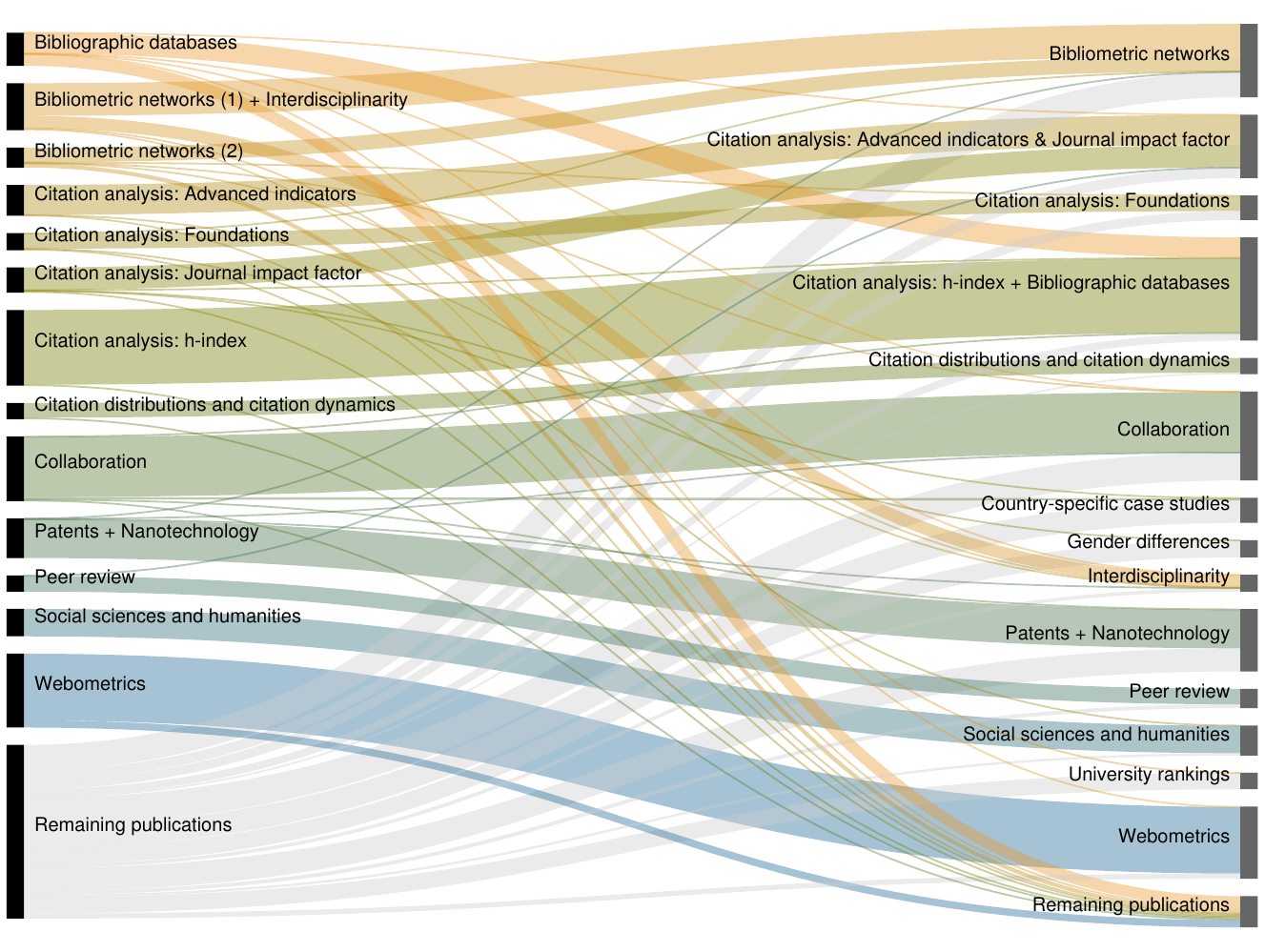} \\
	\caption{{\bf Alluvial diagram of the clusterings obtained by the map equation methods Metimap and Infomap.} The diagram shows the overlap between the largest scientometric clusters returned by Metimap and Infomap on the Library \& Information Science citation network (left and right, respectively). `Remaining publications' are included in one of the clusters in the Metimap (Infomap) clustering but not included in any of the clusters in the Infomap (Metimap) clustering. See~\tblref{expert} for details of the clusterings.}
	\label{fig:expert}
\end{figure}

The clusterings that we are most satisfied with are obtained by Metimap and Infomap. In~\tblref{expert}, we present for each of these clusterings a list of all scientometric clusters with at least $50$ publications. For each cluster, we report the number of publications included in the cluster or equivalently the cluster size $s$ and we provide an indication of the topic that is represented by the cluster. \figref{expert} compares the Metimap and Infomap clusterings by showing the overlap of scientometric clusters using an alluvial diagram.

\begin{table}[!h] \begin{adjustwidth}{-2.25in}{0in}
	\caption{{\bf Statistics of the clusterings obtained by the map equation methods Metimap and Infomap.} The methods are applied to the Library \& Information Science citation network and the largest scientometric clusters with $s\ge 50$ are shown. See~\figref{expert} for a comparison of the clusterings and text for the interpretation.}
	\begin{tabular}{llr} \hline
		{\bf Method} & {\bf Topic} & {\bf Size $s$}  \\ \hline
                	Metimap & Citation analysis: h-index & $262$ \\
		& Webometrics & $256$ \\
		& Collaboration & $224$ \\
		& Bibliometric networks (1) + Interdisciplinarity & $163$ \\
		& Patents + Nanotechnology & $137$ \\
		& Bibliographic databases & $115$ \\
		& Citation analysis: Advanced indicators & $107$ \\
		& Social sciences and humanities & $95$ \\
		& Citation analysis: Journal impact factor & $87$ \\
		& Bibliometric networks (2) & $69$ \\
		& Citation analysis: Foundations & $59$ \\
		& Citation distributions and citation dynamics & $56$ \\
		& Peer review & $56$ \\ \hline
		Infomap & Citation analysis: h-index + Bibliographic databases & $358$ \\
		& Collaboration & $308$ \\
		& Bibliometric networks & $254$ \\
		& Webometrics & $250$ \\
		& Citation analysis: Advanced indicators \& Journal impact factor & $220$ \\
		& Patents + Nanotechnology & $216$ \\
		& Social sciences and humanities & $104$ \\
		& Country-specific case studies & $87$ \\
		& Citation analysis: Foundations & $85$ \\
		& Peer review & $67$ \\
		& Gender differences & $59$ \\
		& Interdisciplinarity & $59$ \\
		& University rankings & $57$ \\
		& Citation distributions and citation dynamics & $56$ \\ \hline
	\end{tabular}
	\label{tbl:expert}
\end{adjustwidth} \end{table}

Metimap and Infomap both offer a reasonable perspective on the main topics in the field of scientometrics. As can be seen in~\tblref{expert}, the clustering returned by Metimap has a somewhat higher resolution than that of Infomap and consequently some topics that are covered by a single cluster in the case of Infomap are distributed over multiple clusters in the case of Metimap. We have a slight preference for Infomap over Metimap because the way in which topics are distributed over multiple clusters in the case of Metimap does not always seem fully satisfactory to us. For instance, we prefer to have a single cluster covering the topic of bibliometric networks instead of the two clusters that are provided by Metimap. However, we emphasize that the differences between the two clusterings are small and that we have only a weak preference for Infomap. Furthermore, even though Metimap and Infomap gave the best clusterings obtained in our study, it should be mentioned that these clusterings sometimes suffer from questionable assignments of publications to clusters. This is a problem especially for smaller clusters. In the case of clusters with fewer than $100$ publications, we often observe that a significant share of the publications assigned to a cluster (e.g. about $25\%$ of the publications) are only weakly related to the main topic of the cluster.

In the case of the clusterings obtained by Metimap and Infomap, we also investigated the effect of applying our post-processing approach (see~\metref). Due to the relatively small size of the Library \& Information Science citation network, the effect of the post-processing approach on the main clusters obtained in the Metimap and Infomap clusterings is small. The number of publications that are reassigned from small clusters to larger clusters, i.e. clusters with at least $50$ publications, is very limited. Given the small effect of the post-processing approach, no significant influence on the quality of the clusters could be observed.

\paragraph{Large-scale clustering analysis.} In the following, we analyze the large-scale behavior of different clustering methods. We limit the analysis to the Louvain modularity optimization method, the map equation algorithm Metimap, the label propagation algorithm BPA and the spectral analysis approach Metilus. These were selected since they can cluster the All Fields citation network in about an hour. %(see~\tblref{wos}).

\tblref{posts} shows bibliometric statistics of the clusterings obtained by the selected methods applied to the Physics citation network (see~\tblref{nets}). Compared to the clusterings obtained for the Library \& Information Science network in~\tblref{bibs}, one can observe a notable increase in the average cluster size $S$ and the effective orders of magnitude $O_5$. The clusterings thus include at least some much larger clusters. Yet, the effective diameter $D_{90}$ and the clustering coverage $K/k$ remain comparable. The clusterings returned by modularity optimization and label propagation methods (i.e. Louvain and BPA) again cover around $80\%$ of the links, while the spectral method Metimap gives $K/k$ below $40\%$. Finally, despite a substantial increase in the network size, the method uncertainty $U$ stays about the same, while the complexity $T$ obviously increases.

\begin{table}[!h] \begin{adjustwidth}{-2.25in}{0in}
	\caption{{\bf Bibliometric statistics of the clusterings obtained by selected methods.} The methods are applied to Physics citation network and bibliometric statistics of the clusterings with and without post-processing are shown. See~\metref for the definitions of statistics and the details of clustering post-processing approach.}
	\begin{tabular}{lrrrrrr} \hline
		{\bf Method} & {\bf Size $S$}& {\bf Orders $O_{5}$} & {\bf Diameter $D_{90}$} & {\bf Coverage $K/k$} & {\bf Uncertainty $U$} &  {\bf Complexity $T$}  \\ \hline
               	% Graclus & $493.4$ & $3.14$ & $7.90$ & $58.7\%$ & $0.402$ & $320.7$ sec \\
                	Louvain & $169.5$ & $4.62$ & $9.88$ & $88.3\%$ & $0.172$ & $89.8$ sec \\
		Metilus & $50.0$ & $2.29$ & $4.53$ & $37.5\%$ & $0.330$ & $140.7$ sec \\ 
		BPA & $43.5$ & $4.58$ & $5.36$ & $76.7\%$ & $0.212$ & $276.0$ sec \\
               	Metimap & $26.5$ & $3.28$ & $3.68$ & $58.8\%$ & $0.122$ & $459.5$ sec \\ \hline
               	% Graclus+post. & $250.4$ & $2.47$ & $6.85$ & $56.2\%$ & $0.402$ & $415.7$ sec \\
                	Louvain+post. & $147.5$ & $3.70$ & $6.92$ & $73.1\%$ & $0.238$ & $134.9$ sec \\
                	Metilus+post. & $51.3$ & $2.23$ & $4.69$ & $37.4\%$ & $0.331$ & $144.7$ sec \\ 
               	BPA+post. & $72.6$ & $4.56$ & $5.39$ & $74.9\%$ & $0.217$ & $340.8$ sec \\
               	Metimap+post. & $44.1$ & $3.29$ & $4.28$ & $59.0\%$ & $0.148$ & $500.3$ sec \\ \hline
	\end{tabular}
	\label{tbl:posts}
\end{adjustwidth} \end{table}

\tblref{posts} also shows the effect of the clustering post-processing approach presented in~\metref that first tries to further partition the largest clusters with $s>s_{giant}$ and then merges the tiny clusters with larger ones for $s<s_{tiny}$, $s_{tiny}=15$ and $s_{giant}=10^4$. In the case of the map equation, label propagation and spectral methods (i.e. Metimap, Metilus and BPA), the post-processing approach has no apparent affect on the largest clusters. Due to the merging of tiny clusters, the average cluster size $S$ increases, while all the remaining statistics remain roughly the same (see~\tblref{posts}). On the other hand, the post-processing manages to further partition the largest clusters returned by the modularity optimization method Louvain. This decreases the cluster size $S$, and also the effective orders $O_5$ and the effective diameter $D_{90}$. However, the clustering coverage $K/k$ decreases as well, while the method uncertainty $U$ increases (see~\tblref{posts}).

\begin{figure}[!h] 
	\includegraphics[width=1.0\textwidth]{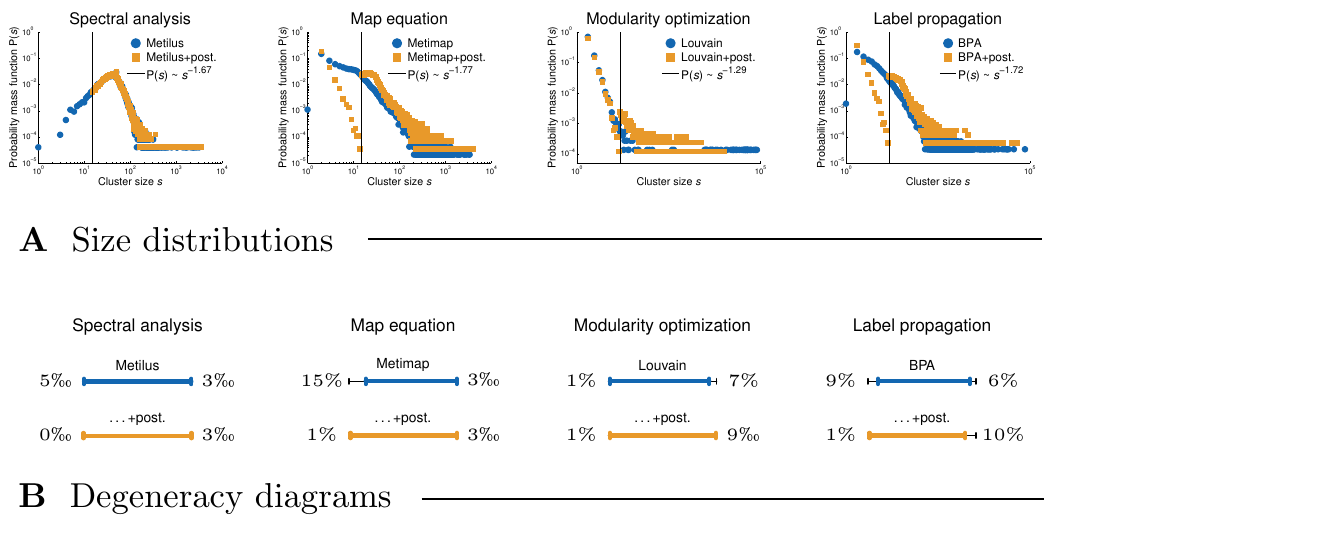} \\
%	\includegraphics[width=0.17\textwidth]{physics_2004_2013_post_size_specter} ~~
%	\includegraphics[width=0.17\textwidth]{physics_2004_2013_post_size_map} ~~
%	\includegraphics[width=0.17\textwidth]{physics_2004_2013_post_size_Q} ~~
%	\includegraphics[width=0.17\textwidth]{physics_2004_2013_post_size_label} \\[7pt]
%	\tikz{\node{{\bf A}~~Size distributions}; \draw[thick] (1.95,0) -- (8.8,0);} \\~\\
%	% \includegraphics[width=0.17\textwidth]{physics_2004_2013_post_degenerate_specter} ~~
%	% \includegraphics[width=0.17\textwidth]{physics_2004_2013_post_degenerate_map} ~~
%	% \includegraphics[width=0.17\textwidth]{physics_2004_2013_post_degenerate_Q} ~~
%	% \includegraphics[width=0.17\textwidth]{physics_2004_2013_post_degenerate_label} \\[7pt]
%	\begin{tabular}{cccc}
%		\dgd{Spectral analysis}{\dgdt{0.00547002047761649}{0.99717601832771}{5\permil}{3\permil}{Metilus}}{\dgdb{0.000025536220088168868}{0.9974350285600329}{0\permil}{3\permil}{\dots+post.}} & 
%		\dgd{Map equation}{\dgdt{0.15443373634622898}{0.9965858479078945}{15\%}{3\permil}{Metimap}}{\dgdb{0.014631848773693964}{0.9971533194654094}{1\%}{3\permil}{\dots+post.}} & 
%		\dgd{Modularity optimization}{\dgdt{0.014863296101794668}{0.9319994779261671}{1\%}{7\%}{Louvain}}{\dgdb{0.014634280794654743}{0.9912135136055359}{1\%}{9\permil}{\dots+post.}} & 
%		\dgd{Label propagation}{\dgdt{0.09434011975271211}{0.9394187632038471}{9\%}{6\%}{BPA}}{\dgdb{0.014634280794654743}{0.8979896914738209}{1\%}{10\%}{\dots+post.}}
%	\end{tabular} \\[1pt]
%	\tikz{\node{{\bf B}~~Degeneracy diagrams}; \draw[thick] (2.2,0) -- (8.525,0);} \\
	\caption{{\bf Size distributions and degeneracy of the clusterings obtained by the selected methods.} The methods with and without post-processing are applied to the Physics citation network, while the panels~{\bf A} and~{\bf B} show cluster size distributions $\mathrm{P}(s)$ and clustering degeneracy diagrams $D$, respectively. Vertical lines in panel~{\bf A} represent the threshold size $s_{tiny}=15$. See text for the definition of clustering degeneracy and~\metref for the details of the clustering post-processing approach.}
	\label{fig:posts}
\end{figure}

\figref{posts} shows the impact of the post-processing approach on the cluster size distributions $\mathrm{P}(s)$ and the clustering degeneracy diagrams $D$. All distributions $\mathrm{P}(s)$ remain conceptually the same, with the difference that most tiny clusters have been merged with larger ones (see~\figref{posts}, panel~{\bf A}). Notice that a small number of tiny clusters with $s<15$ remain, which correspond to disconnected components that could obviously not be merged with other clusters (see~\tblref{nets} for the size of LCC). Still, the degeneracy diagrams $D$ show that post-processing effectively removes tiny clusters, and also the giant cluster in the case of the modularity optimization method Louvain, but fails to further partition the giant cluster in the case of the label propagation algorithm BPA (see right-hand side of~\figref{posts}, panel~{\bf B}).

Last, we apply the selected methods to the All Fields citation network (see~\tblref{nets}). \tblref{wos} shows different statistics of the obtained clusterings. Compared to those obtained for the Physics citation network in~\tblref{posts}, we can again observe an increase in the average cluster size $S$ and the effective orders $O_5$. Thus the size of the largest clusters further increases. Yet, as before, the clustering coverage $K/k$ of different methods remains roughly the same, while the differences between the methods can also clearly be observed in the average internal degree $K$. \tblref{wos} also shows the statistics of the clusterings after the post-processing approach, which has exactly the same effect on the clusterings as in~\tblref{posts}. Notice also that the post-processing does not substantially increase the running time of the methods.

\begin{table}[!h] \begin{adjustwidth}{-2.25in}{0in}
	\caption{{\bf Statistics of the clusterings obtained by the selected methods.} The methods are applied to the All Fields citation network and different statistics of the clusterings with and without post-processing are shown. See~\metref for the definitions of the statistics and the details of the clustering post-processing approach.}
	\begin{tabular}{lrrrrrr} \hline
		{\bf Method} & {\bf Size $S$}& {\bf Orders $O_{5}$} & {\bf Degree $K$} & {\bf Coverage $K/k$} & {\bf Flake $F$} &  {\bf Complexity $T$}  \\ \hline
               	Louvain & $334.4$ & $5.74$ & $18.53$ & $83.9\%$ & $5.3\%$ & $52.1$ min \\
               	BPA & $105.4$ & $6.22$ & $18.50$ & $83.8\%$ & $7.2\%$ & $66.2$ min \\
		Metilus & $50.0$ & $2.33$ & $5.91$ & $26.8\%$ & $68.9\%$ & $30.0$ min \\
               	Metimap & $33.2$ & $3.55$ & $10.30$ & $46.6\%$ & $45.0\%$ & $94.2$ min \\ \hline
               	Louvain+post. & $320.9$ & $4.88$ & $15.20$ & $68.8\%$ & $17.1\%$ & $78.9$ min \\
               	BPA+post. & $167.1$ & $6.20$ & $18.04$ & $81.7\%$ & $9.0\%$ & $114.3$ min \\
               	Metilus+post. & $51.5$ & $2.24$ & $5.92$ & $26.8\%$ & $68.9\%$ & $34.3$ min \\
               	Metimap+post. & $58.9$ & $3.55$ & $10.33$ & $46.8\%$ & $44.5\%$ & $98.9$ min \\ \hline
	\end{tabular}
	\label{tbl:wos}
\end{adjustwidth} \end{table}

\begin{figure}[!h] 
	\includegraphics[width=1.0\textwidth]{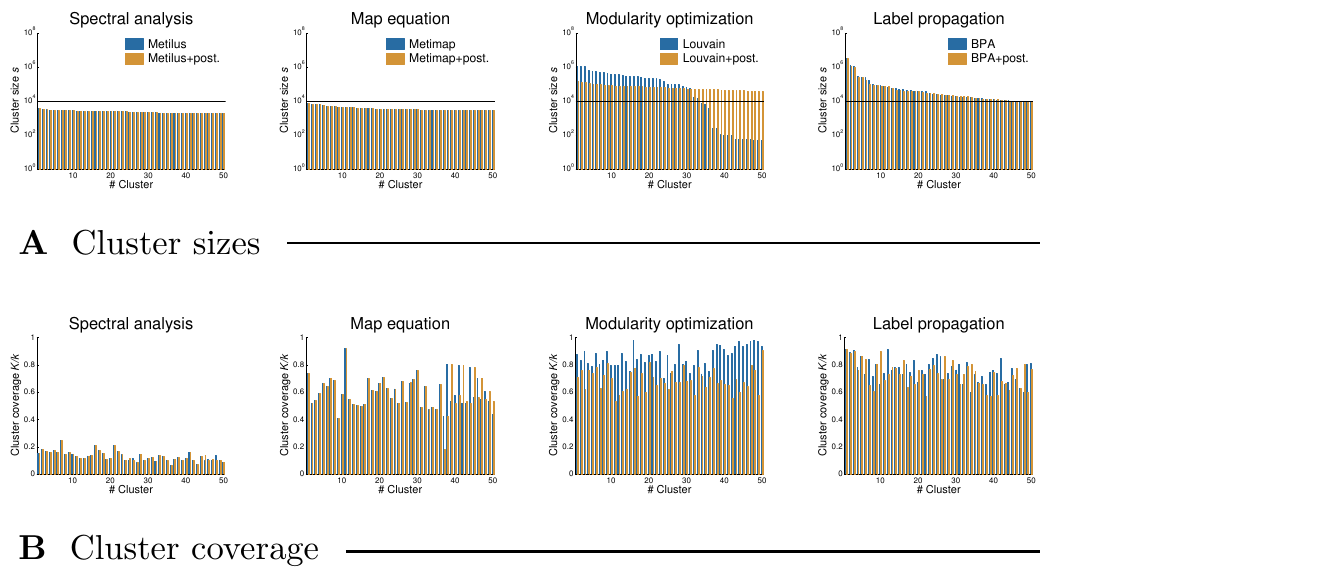} \\
%	\includegraphics[width=0.17\textwidth]{wos_2004_2013_specter_s} ~~
%	\includegraphics[width=0.17\textwidth]{wos_2004_2013_map_s} ~~
%	\includegraphics[width=0.17\textwidth]{wos_2004_2013_Q_s} ~~
%	\includegraphics[width=0.17\textwidth]{wos_2004_2013_label_s} \\[7pt]
%	\tikz{\node{{\bf A}~~Cluster sizes}; \draw[thick] (1.5,0) -- (9.15,0);} \\~\\
%	\includegraphics[width=0.17\textwidth]{wos_2004_2013_specter_k} ~~
%	\includegraphics[width=0.17\textwidth]{wos_2004_2013_map_k} ~~
%	\includegraphics[width=0.17\textwidth]{wos_2004_2013_Q_k} ~~
%	\includegraphics[width=0.17\textwidth]{wos_2004_2013_label_k} \\[7pt]
%	\tikz{\node{{\bf B}~~Cluster coverage}; \draw[thick] (1.8,0) -- (8.85,0);} \\~\\
	\caption{{\bf Sizes and coverage of the largest clusters obtained by the selected methods.} The methods with and without post-processing are applied to the All Fields citation network, while the panels~{\bf A} and~{\bf B} show the sizes $s$ and coverage $K/k$ of the largest $50$ clusters, respectively. Horizontal lines in panel~{\bf A} represent the threshold size $s_{giant}=10^4$. See text for the definition of cluster coverage.}
	\label{fig:wos}
\end{figure}

To better understand the nature of different clusterings and the effects of the post-processing approach, \figref{wos} shows the sizes $s$ and coverage $K/k$ of the largest $50$ clusters returned by the selected methods (see~\metref). The coverage $K/k$ of an individual cluster is defined as the average internal degree of the nodes in the cluster divided by the total degree of these nodes. As already lengthly discussed above, the spectral analysis approach Metilus returns clusters with very low $K/k\approx 15\%$ (see left-hand side of~\figref{wos}, panel~{\bf B}), while the modularity optimization and label propagation methods (i.e. Louvain and BPA) give clusters with very high $K/k\approx 80\%$ (see right-hand side of~\figref{wos}, panel~{\bf B}). For the map equation algorithm Metimap, $K/k\approx 60\%$. One can also observe that, in the case of the label propagation algorithm BPA, the post-processing approach fails to further partition the largest clusters with $s>s_{giant}$, where $s_{giant}$ is represented by horizontal lines in~\figref{wos}, panel~{\bf A}. On the contrary, the post-processing does partition the largest clusters in the case of the modularity optimization method Louvain. However, the results are far from satisfactory. Each cluster with $s>s_{giant}$ is indeed split into smaller clusters, but the number of such clusters thus actually increases (see middle of~\figref{wos}, panel~{\bf A}).

% % % % % % % % % % % % % % % % % % % % % % % % % % 
%
%					DISCUSSION
%
% % % % % % % % % % % % % % % % % % % % % % % % % %

\section*{Discussion} 

Which methods for graph partitioning and community detection perform best for the purpose of grouping scientific publications into clusters? In this paper, we have carried out an extensive analysis comparing the performance of a large number of methods. The methods have been applied to a number of networks of publications connected by direct citation relations. We have studied the statistical properties of the results provided by the different methods, and we have also performed an expert-based assessment of the results.

From a bibliometric point of view, a good clustering of publications ideally should have a number of properties. First of all, although it is natural to expect that there will be larger and smaller clusters, it is inconvenient for practical purposes if there are very large differences in the size of clusters. As a rule of thumb, we ideally would like the difference in size between the largest and the smallest clusters to be no more than an order of magnitude. Second, if it turns out to be inevitable that some publications end up in very small clusters, for instance because these publications have almost no citation relations with other publications, then at least we would prefer the number of publications assigned to these insignificant clusters to be as limited as possible. Third, we would like the results of a clustering method to be reasonably stable. Many methods include a random element, in which case different runs of a method may yield different results. However, running the same method multiple times should not affect the results too much, and the results should also be reasonably robust to small changes in a citation network of publications. Fourth, the computing time of a clustering method should not be excessive. This is especially important when one aims to apply a method to networks consisting of large numbers of publications and citation relations. Finally, and perhaps most importantly, the results produced by a clustering method should make intuitive sense. Experts should be able to recognize the scientific topics represented by clusters of publications.

Our analysis shows that most clustering methods yield results with large differences in the size of clusters. The larger clusters are typically several orders of magnitude larger than the smaller clusters. Sometimes more than half of the publications in a citation network are all assigned to the same cluster. This was for instance observed for the results obtained from the Links and SCP methods in the Library \& Information Science citation network. The only methods that yield clusters of more or less similar size are the spectral methods (e.g. Graclus). These methods produce results that are characterized by a much more uniform cluster size distribution. % Depending on the cluster size distribution, there can be large differences in the share of all citation relations that are covered by clusters. 
Depending on the cluster size distribution and also on the resolution of a clustering, there can be large differences in the share of all citation relations that are covered by clusters. Coverage for instance ranges from less than $30\%$ to more than $85\%$ in the Library \& Information Science citation network. Clustering methods also often assign a significant share of the publications in a citation network to very small clusters. In the Library \& Information Science citation network, the Graclus and Infomap methods for instance assign more than $25\%$ of the publications to clusters consisting of fewer than $15$ publications. The stability or robustness of the results obtained from a clustering method also partly depends on the size of the clusters produced by the method. Not surprisingly, methods that produce one or more very large clusters tend to yield relatively robust results. Furthermore, in the Library \& Information Science citation network, spectral and statistical methods (e.g. Graclus and OSLOM) produce results with a relatively low robustness, while Infomap and modularity optimization yield quite robust results.

In terms of computing time, there are substantial differences between the various methods. For instance, clustering the publications in the Library \& Information Science citation network takes more than $100$ times longer for the slowest method than for the fastest method. Modularity optimization methods (e.g. Louvain), label propagation (e.g. BPA), and spectral analysis methods (e.g. Graclus) perform best in terms of computing time. Other methods require a more significant amount of computing time, making them less suitable for applications on large citation networks.

Turning now to the expert-based assessment of the results produced by different clustering methods for the scientometrics subfield within the Library \& Information Science citation network, we find that the Infomap and Metimap (i.e. Infomap combined with spectral method METIS) methods give the most satisfactory results, with a slight preference for the Infomap results over the results obtained from Metimap. Other methods, such as OSLOM and Louvain, provide less satisfactory results.

Our analysis seems to provide most support for the use of Infomap and related methods such as Metimap to cluster the publications in a citation network. Infomap has the best performance in our expert-based assessment, and it yields quite robust results. Compared with some of the other methods, Infomap has a relatively high computing time, but this can be overcome by using Metimap in larger citation networks. The price that we pay for the good performance of Infomap seems to be the assignment of a relatively large number of publications to small clusters. Paying this price seems necessary to obtain high-quality clustering results. In large citation networks, a post-processing procedure can be applied to minimize the number of small clusters, but the effect of the use of such a procedure on the quality of the clustering results is not clear.

The promising results obtained for Infomap are in line with earlier findings reported in the network science literature~\cite{LF09b}. Although Infomap has been introduced in the bibliometric literature~\cite{BELR14} and has been applied to citation networks in a number of studies~\cite{RB08,RB10,RB11b,MBAR13}, the method has not yet gained a widespread popularity in the bibliometric community, where researchers seem to prefer the use of modularity-based methods. Our findings suggest that the bibliometric community could benefit from exploring the use of other clustering methods in addition to modularity-based methods. Infomap seems to be of particular interest. Future studies should reveal whether Infomap indeed consistently performs well in applications to citation networks.

\paragraph{Limitations of the analysis.} It is important to emphasize that our results should be interpreted cautiously because of a number of limitations of our analysis. One obvious limitation is that, despite the large number of clustering methods included in our analysis, we did not exhaustively cover all methods proposed in the literature. The selection of the methods included in our analysis was made based on the popularity of a method and to some degree also on our familiarity with a method. In addition, the availability of source code played a role as well. Many methods discussed in the literature are not included in our analysis. In particular, methods that produce overlapping clusters~\cite{BKN11,GB13} or clusters at multiple levels of resolution~\cite{RN09,TKV13a} are not covered. Also, we for instance do not cover some recently developed principled methods based on statistical inference~\cite{Pei15}.

A second limitation is that each clustering method was applied using the default parameter settings. We did not try to optimize the parameter values of the different methods. So the performance of some methods may have been better if we had used optimized parameter values for these methods. Some methods for instance have a parameter that can be used to fine-tune the level of granularity of the clustering results. One could use such a parameter to try to obtain results at similar levels of granularity for different methods, and in that way a more accurate comparison between different methods may be possible. We did not explore this possibility in our analysis, but we do consider this an interesting direction for future research. We note that the clustering method proposed by two of us in an earlier paper~\cite{WV12} requires a careful choice of parameter values. For this reason, this method was not included in our present analysis.

A third limitation is our exclusive focus on undirected and unweighted networks of direct citation relations between publications. We did not consider the possibility of taking into account the direction of a citation relation, and we did not test the effect of assigning weights to citation relations~\cite{WV12}. We also did not study the use of indirect citation relations between publications, in particular co-citation and bibliographic coupling relations.

Finally, we should emphasize the limitations of our expert-based assessment of the clustering results obtained for the scientometrics subfield within the Library \& Information Science citation network. The expert-based assessment was carried out at a high level of detail by two experts with an extensive expertise in the field of scientometrics. Nevertheless, any expert-based assessment will necessarily be of a subjective nature, and different experts therefore may not always reach the same conclusions. Moreover, experts typically have a deep understanding of the literature only in a relatively small area of science. This for instance explains why in our expert-based assessment we could not cover the entire field of library and information science but only the subfield of scientometrics. Unfortunately, it is difficult to say to what extent conclusions reached for such a relatively small area of science can be expected to generalize to other areas. For this reason, the findings of our expert-based assessment should be interpreted with some caution.

% % % % % % % % % % % % % % % % % % % % % % % % % % 
%
%					SUPPLEMENT
%
% % % % % % % % % % % % % % % % % % % % % % % % % %

\section*{Acknowledgments}

We thank numerous authors for kindly providing the source code of their methods. This work has been supported in part by the Slovenian Research Agency Program No.\ P2-0359.

%\section*{Author Contributions}
%
%Conceived and designed the experiments: L\v{S} NJvE LW. Performed the experiments: L\v{S}. Analyzed the data: L\v{S} NJvE LW. Contributed reagents/materials/analysis tools: NJvE. Wrote the paper: L\v{S} LW.

%\nolinenumbers

%\bibliography{bibliography}

\end{document}